# On systems having Poincaré and Galileo symmetry

Peter Holland

Green Templeton College
University of Oxford
Oxford OX2 6HG
England

peter.holland@gtc.ox.ac.uk

**Abstract**

Using the wave equation in $d \geq 1$ space dimensions it is illustrated how dynamical equations may be simultaneously Poincaré and Galileo covariant with respect to different sets of independent variables. This provides a method to obtain dynamics-dependent representations of the kinematical symmetries. When the field is a displacement function both symmetries have a physical interpretation. For $d = 1$ the Lorentz structure is utilized to reveal hitherto unnoticed features of the non-relativistic Chaplygin gas including a relativistic structure with a limiting case that exhibits the Carroll group, and field-dependent symmetries and associated Noether charges. The Lorentz transformations of the potentials naturally associated with the Chaplygin system are given. These results prompt the search for further symmetries and it is shown that the Chaplygin equations support a nonlinear superposition principle. A known spacetime mixing symmetry is shown to decompose into label-time and superposition symmetries. It is shown that a quantum mechanical system in a stationary state behaves as a Chaplygin gas. The extension to $d > 1$ is used to illustrate how the physical significance of the dual symmetries is contingent on the context by showing that Maxwell's equations exhibit an exact Galileo covariant formulation where Lorentz and gauge transformations are represented by field-dependent symmetries. A natural conceptual and formal framework is provided by the Lagrangian and Eulerian pictures of continuum mechanics.

## 1 Introduction

That a system of equations could admit both Lorentz and Galileo boosts as exact symmetries seems nonsensical at first sight. The relation between the two is customarily presented as one of asymptotic inclusion, the Galilean transformation emerging from the Lorentzian in the limit of low velocities relative to the speed of light. Nevertheless, it is well known that a dynamical equation can be both Poincaré and Galileo covariant if, for example, it possesses the general linear group as a symmetry (of which the Poincaré and Galileo groups are subgroups; for an example see [1]). Here we examine how a physical system may be simultaneously 'relativistic' and 'non-relativistic' by virtue of admitting descriptions in terms of two sets of independent variables, its governing equations being Poincaré covariant with respect to one set and Galileo covariant with respect to the other. Since the (Lorentz or Galileo) boost symmetry pertaining to one set of variables has a realization in the other set, each set of variables exhibits two independent symmetries involving velocity parameters. This approach then provides a method to discover alternative representations of the kinematical symmetries, which may become dynamical by virtue of depending on solutions of the dynamical equations. The extent to which the two symmetries describe operations involving physical observers, or have a more arcane character, depends on the meaning afforded to the variables chosen to describe the system that are subject to the relevant transformations and the nature of the governing equations. Obviously, for a given set of equations at most one of the velocity parameters may characterize a spatial boost, i.e., a transformation between inertial reference frames in uniform relative motion. The other parameter must then relate to a different notion of 'boost'.

Although it does not seem to be widely known, this dual covariance is a feature of field equations used in physics and is exemplified by the wave equation in $d \geq 1$ space dimensions, which will be our focus. When the field variable is a displacement function this provides an example where both symmetries have a physical interpretation. We shall examine the $d = 1$ case in detail as this already illustrates the key symmetry properties, which persist in higher dimensions but take more complex forms. The cases we consider are contained in a class of theories where each set of independent variables spans a $(d,1)$-dimensional coordinate-time space, with a common time coordinate. A natural conceptual and formal framework is provided by continuum mechanics, in which context the sets of envisaged variables correspond to the Lagrangian and Eulerian pictures. The variables are linked by the standard transformation rules connecting the two descriptions, where fields in one picture become independent variables in the other.

In elementary treatments of the derivation of continuous field equations from discrete mechanical models, the wave equation is obtained as the limit of the Newtonian dynamics of a linear chain of coupled harmonic oscillators [2]. Denoting the limiting particle label by $a \in \mathbb{R}$ (having the dimension of length) and assuming the limiting mass density $k$ of the oscillators in the reference state is uniform, one obtains for the longitudinal displacement $q(a,t)$ of the $a$th particle in a one-dimensional chain (assumed infinite) at time $t$:

$$\frac{\partial^2 q}{\partial t^2} - c^2 \frac{\partial^2 q}{\partial a^2} = 0 \qquad (1.1)$$

where $c = \sqrt{Y/k}$ and $Y$ is Young's modulus. Note that this equation holds for finite amplitude $q$. The description is a *Lagrangian*, or *material*, one [3], which specifies the state of a system by the motion of each of the mass points (independent variables $a,t$; state variable position $q(a,t)$). The description is completed by the law of mass conservation whose solution gives the density at time $t$: $\rho(q(a,t),t) = k(\partial q/\partial a)^{-1}$. Following the usual approach of continuum mechanics, we may consider an alternative description, the *Eulerian*, or *spatial*, one, which specifies the state at a fixed spacetime point (independent variables $x = q(a,t),t$; state variables density $\rho(x,t)$ and velocity $v(x,t)$). As we shall see, the Lagrangian and Eulerian formulations each exhibit relativistic and non-relativistic symmetries. For this system, the covariance of the field equations in the two pictures with respect to a physical boost is described by Galileo's transformation, not Lorentz's. The latter describes a time-dependent label-time substitution.

It was observed by Earnshaw [4] in 1860 that the Eulerian equation of state corresponding to the process (1.1) is $p(\rho) = A - 2\lambda/\rho$, $\lambda = \tfrac{1}{2}c^2 k^2 > 0$ (for a review see [5]). Several authors [4,6,7] have expressed disquiet at applying this equation of state to realistic models of known substances. Thus, it cannot approximate Boyle's law $p \propto \rho$ and it implies negative pressure. Nevertheless, following Chaplygin this equation of state has been found useful in various contexts in $d \geq 1$ dimensions, such as modelling unstable flows [8], and a medium in such a state is often referred to as a 'Chaplygin gas', a usage we shall follow. As noted, in one dimension the Chaplygin equation of state represents an elastic string, which thereby provides pedagogical insight. In fact, a general class of steady 1+1 fluid flows reduces to a Chaplygin flow, including that associated with quantum mechanical stationary states, as we shall see later.

The connection between the $d = 1$ wave equation and the Chaplygin equations has been noted in the context of the Nambu-Goto brane theory [9,10]. In fact, in that setting the Galileo covariant Chaplygin equations are revealed to be rich in relativistic allusions. Thus, in $d \geq 1$ dimensions they can be mapped into the Poincaré covariant Born-Infeld system (by a different change of variables to that considered here) [9]. Intriguingly, the Chaplygin gas is also a non-relativistic limit of the Born-Infeld model [9]. An example of the dual symmetries operating in different dimensions is also known: it was discovered by Bazeia and Jackiw [9,11,12] that the $d$-dimensional Chaplygin gas exhibits a spacetime mixing symmetry that is a realization of the Poincaré group in $d+1$ dimensions. This symmetry is 'field-dependent' in that the coordinate change involves the dynamical fields that are subject to that change.

It is surprising that the connection between the symmetries in the $d = 1$ Lagrangian and Eulerian formulations has not been explored previously and Secs. 2-8 are devoted to this. We exploit the linearized form of the Chaplygin fluid to show that the latter admits field-dependent symmetries in addition to the spacetime mixing symmetry. These transformations can arise from kinematical symmetries; two examples are obtained from the Eulerian images of the Lagrangian-picture Lorentz and special conformal transformations. The Eulerian counterpart of a Lorentz boost in the Lagrangian picture involves a mass density potential-dependent transformation of the time, while the Lagrangian counterpart of a Galileo boost in the Eulerian picture is a linear time-dependent displacement of the field. As noted above, both boosts may be implemented by physical operations. We also detail the non-relativistic limits of the Chaplygin theory, which includes a case where the field-dependent Lorentz group reduces to the Carroll group, and examine other aspects such as the Lorentz transformation of the potentials that are naturally associated with the Chaplygin system. These results prompt the search for further symmetries, and additional field-dependent ones emerge from the observation that the Chaplygin equations support a nonlinear superposition principle induced by the linear method of combining Lagrangian-picture solutions. This result is used to show that the mixing symmetry may be decomposed into two transformations induced by label-time and superposition symmetries of the wave equation.

The physical character of the Poincaré and Galileo symmetries extends to the wave equation in $d > 1$ dimensions when the field is treated as a displacement function (in this case the Eulerian version of the wave equation is not the Chaplygin gas). To illustrate how the physical significance of the dual symmetries is contingent on the context, we show that Maxwell's equations exhibit an exact Galileo covariant formulation alongside the usual Poincaré one, in which Lorentz and gauge transformations are represented by field-dependent symmetries. In that case the covariance of the field equations in the two pictures with respect to a physical boost is described by Lorentz's transformation, not Galileo's. The latter has just a mathematical significance, being a gauge transformation.

## 2 The Chaplygin gas

For a set of particles having a variable label density $\rho_0(a)$ (which is taken to be $k$ in connection with (1.1)), the density and velocity in the Eulerian description are obtained from the Lagrangian variables via the following formulas:

$$\rho(x,t) = \left(\frac{\partial q(a,t)}{\partial a}\right)^{-1} \rho_0(a)\bigg|_{a(x,t)}, \quad v(x,t) = \frac{\partial q(a,t)}{\partial t}\bigg|_{a(x,t)}. \tag{2.1}$$

It is necessary that $\partial q/\partial a > 0$ to avoid the appearance of singularities [13]. Differentiating (2.1) with respect to *t* and using (1.1) we deduce the continuity and Euler equations:

$$\frac{\partial \rho}{\partial t} + \frac{\partial}{\partial x}(\rho v) = 0 \tag{2.2}$$

$$\frac{\partial v}{\partial t} + v\frac{\partial v}{\partial x} = \frac{\partial}{\partial x}\left(\frac{\lambda}{\rho^2}\right) \tag{2.3}$$

where $\lambda = \tfrac{1}{2}c^2 k^2 > 0$. These are the equations governing a one-dimensional Chaplygin gas.

For a one-dimensional fluid, the identity transformation in the Eulerian picture,

$$x' = x, \quad t' = t, \quad \rho'(x',t') = \rho(x,t), \quad v'(x',t') = v(x,t), \tag{2.4}$$

corresponds to the following transformation in the Lagrangian picture:

$$q'(a',t') = q(a,t), \quad t' = t, \quad \rho_0'(a')\frac{\partial a'}{\partial a} = \rho_0(a), \quad \frac{\partial a'}{\partial t} = 0, \tag{2.5}$$

i.e., a time-independent diffeomorphism $a'(a)$ that relabels the particles. We can use this freedom to choose a natural labelling that might be suggested by the physical setting, or that simplifies the Lagrangian problem, without affecting the Eulerian functions. We consider two choices of labelling characterized by the connection with initial position:

(a) *Label as initial position*. In an arbitrary labelling system, the linear Lagrangian-picture equation implied by (2.1) and (2.3) is

$$\frac{\partial^2 q}{\partial t^2} = \frac{2\lambda}{\rho_0(a)^2}\left(\frac{\partial^2 q}{\partial a^2} - \frac{\partial}{\partial a}\log\rho_0(a)\frac{\partial q}{\partial a}\right). \tag{2.6}$$

This equation may be derived directly from the discrete model mentioned in Sec. 1 by allowing the density of particles to vary. Using relativistic notation, with $a^\mu = (ct, a)$ and $\partial_\mu = \partial/\partial a^\mu$, (2.6) may be expressed as a two-dimensional wave equation with a label-dependent metric:

$$\frac{1}{\sqrt{-g}}\partial_\mu\left(\sqrt{-g}g^{\mu\nu}\partial_\nu q\right)=0, \quad g_{\mu\nu}=\begin{pmatrix} 1 & 0 \\ 0 & -\dfrac{c^2\rho_0(a)^2}{2\lambda} \end{pmatrix}, \quad \mu,\nu,=0,1. \quad (2.7)$$

A natural choice for labelling is the initial particle position, $a = q_0(a)$, which implies that the element $da$ coincides with an elementary spatial volume. Evaluating (2.1) at $t = 0$ then gives $\rho(a,0) = \rho_0(a)$, which justifies the notation used for the latter.

(b) *Uniform density*. Although the choice (a) is intuitively appealing, we shall work instead with a labelling that simplifies the Lagrangian equation (2.6). Thus, we transform to the label

$$a' = k^{-1}\int^a \rho_0(a)da \quad (2.8)$$

so that, from (2.5), the density is uniform ($\rho_0'(a') = k$), (2.6) reduces to (1.1) (with $a$ replaced by $a'$), and the metric in (2.7) is transformed globally to Minkowskian form. Evaluating (2.1) at $t = 0$, the label and initial position are connected as follows:

$$q_0(a) = \int^a (k/\rho(a,0))da. \quad (2.9)$$

Of course, if the true initial density is $k$ then $a = q_0(a)$, as in case (a).

The transformations considered in this paper are symmetries of (1.1), i.e., they connect labelling systems for which the density is uniform. They become symmetries of (2.6) upon replacing $a$ by $k^{-1}\int \rho_0(a)da$, and have the property $\rho_0'(a') = \rho_0(a')$.

To employ the wave equation to solve the Chaplygin equations we use as Cauchy data (2.9) for $q_0(a)$ and $\varphi(a) \equiv \partial q/\partial t\big|_{t=0} = v_0(q_0(a))$. Then the general solution of (1.1),

$$q(a,t) = \frac{1}{2}\left[q_0(a-ct) + q_0(a+ct) + (1/c)\int_{a-ct}^{a+ct}\varphi(\tau)d\tau\right], \quad (2.10)$$

generates the general solution of the one-dimensional Chaplygin gas by differentiation and inversion using the formulas (2.1) (with $\rho_0 = k$).

Regarding the variational basis of the Chaplygin theory, the standard Lagrangian generating (1.1) is

$$L_L = \int\frac{1}{2}k\left[\left(\frac{\partial q}{\partial t}\right)^2 - c^2\left(\frac{\partial q}{\partial a}\right)^2\right]da. \quad (2.11)$$

The Lagrangian for (2.6) is obtained from this by replacing $k$ with $\rho_0(a)$ (also in $c = \sqrt{2\lambda}/k$). Starting from the canonical formalism in the Lagrangian picture, a variational formulation in the Eulerian picture may be obtained following a standard procedure [14,15], which links the relativistic and non-relativistic theories by a canonical transformation. The derived Lagrangian in the Eulerian picture is

$$L_E = \int \left[ \theta \frac{\partial \rho}{\partial t} - \frac{1}{2}\rho \left(\frac{\partial \theta}{\partial x}\right)^2 - \frac{\lambda}{\rho} \right] dx \tag{2.12}$$

where $\theta$ is the velocity potential. The Euler-Lagrange equations for $\theta$ and $\rho$ are (2.2) and the Bernoulli equation

$$\frac{\partial \theta}{\partial t} + \frac{1}{2}\left(\frac{\partial \theta}{\partial x}\right)^2 - \frac{\lambda}{\rho^2} = 0 \tag{2.13}$$

from which (2.3) may be derived by differentiation. The expressions (2.11) and (2.12) will be used below in the applications we make of Noether's theorem.

Table 1 lists the potentials used in this paper. Their transformation laws follow from those of the basic Eulerian and Lagrangian variables.

| Potential | Lagrangian picture | Eulerian picture |
|---|---|---|
| Velocity | $\phi(a,t) = \int^{a,t} \frac{\partial q}{\partial t}\frac{\partial q}{\partial a} da + \frac{1}{2}\left[c^2\left(\frac{\partial q}{\partial a}\right)^2 + \left(\frac{\partial q}{\partial t}\right)^2\right] dt$ | $\theta(x,t) = \int^{x,t} v\, dx + \left(-\frac{1}{2}v^2 + \lambda/\rho^2\right) dt$ |
| Mass density | $ka$ | $m(x,t) = \int^{x,t} \rho\, dx - \rho v\, dt$ |
| Momentum density | $k\alpha(a,t) = k\int^{a,t} \frac{\partial q}{\partial t} da + c^2 \frac{\partial q}{\partial a} dt$ | $\xi(x,t) = \int^{x,t} \rho v\, dx + 2\rho \frac{\partial \theta}{\partial t} dt$ |
| Energy density | $k\beta(a,t) = k\int^{a,t} \frac{\partial \phi}{\partial t} da + c^2 \frac{\partial \phi}{\partial a} dt$ | $\chi(x,t) = \int^{x,t} \mathfrak{H}\, dx + \rho v \frac{\partial \theta}{\partial t} dt$, $\mathfrak{H} = \frac{1}{2}\rho v^2 + \lambda/\rho$ |

**Table 1**. The 1+1 potentials used in the text. The Eulerian forms are obtained from the Lagrangian ones via formulas such as $\theta(x,t) = \phi(a(x,t),t)$. Each potential exists by virtue of a continuity (or zero *at*- or *xt*-vorticity) equation implied by the wave or Chaplygin equations (such as $\partial(\partial\theta/\partial x)/\partial t = \partial(\partial\theta/\partial t)/\partial x$).

### 3 Poincaré symmetry

#### *3.1 Label-time translation*

Eq. (1.1) is covariant with respect to the substitution

$$q'(a',t') = q(a,t), \quad t' = t + t_0, \quad a' = a + a_0. \tag{3.1}$$

The implied transformation of the Eulerian variables is

$$x' = x, \quad t' = t + t_0, \quad \rho'(x',t') = \rho(x,t), \quad v'(x',t') = v(x,t) \tag{3.2}$$

The Noether currents implied by (3.1) are given by the conserved energy-momentum tensor for the Lagrangian field:

$$t_{\mu\nu} = kc^2 \left( \partial_\mu q \partial_\nu q - \tfrac{1}{2} \eta_{\mu\nu} \partial^\sigma q \partial_\sigma q \right), \quad \eta_{\mu\nu} = \begin{pmatrix} 1 & 0 \\ 0 & -1 \end{pmatrix}, \quad \partial^\mu t_{\mu\nu} = 0. \tag{3.3}$$

The conserved charges are therefore

$$E = \int t_{00}\, da = k \int \frac{1}{2}\left( \left(\frac{\partial q}{\partial t}\right)^2 + c^2 \left(\frac{\partial q}{\partial a}\right)^2 \right) da \tag{3.4}$$

$$C = \int t_{01}\, da = \sqrt{2\lambda} \int \frac{\partial q}{\partial t} \frac{\partial q}{\partial a}\, da, \tag{3.5}$$

which when expressed in terms of the Eulerian variables become

$$E = \int \mathfrak{H}\, dx, \quad \mathfrak{H} = \tfrac{1}{2}\rho v^2 + \lambda/\rho \quad \text{(time translation)} \tag{3.6}$$

$$C = \sqrt{2\lambda} \int v\, dx. \quad \text{(relabelling)} \tag{3.7}$$

The energy $E$ coincides with the usual Eulerian expression obtained by direct application of Noether's theorem to the transformation (3.2). In contrast, $C$ does not give the corresponding momentum (i.e., (4.3), associated with a space translation). Rather, $C$ corresponds to a pure label transformation. This is an example where the Noether charge in the *at* variables implies a corresponding charge in the Eulerian variables that differs from the charge obtained by direct application of Noether's theorem in the *xt* variables using the Eulerian equivalent of the Lagrangian symmetry (here the identity) [16].

### *3.2 Label-time Lorentz transformation*

Eq. (1.1) is covariant with respect to a label-time Lorentz transformation:

$$q'(a',t') = q(a,t), \quad t' = \gamma\left(t - \frac{ua}{c^2}\right), \quad a' = \gamma(a - ut), \tag{3.8}$$

where $\gamma = (1 - u^2/c^2)^{-1/2}$. Thus, $\partial_\mu q$ is a covariant Lorentz ('deformation') vector:

$$\frac{\partial q'}{\partial t'} = \gamma\left( \frac{\partial q}{\partial t} + u\frac{\partial q}{\partial a} \right), \quad \frac{\partial q'}{\partial a'} = \gamma\left( \frac{\partial q}{\partial a} + \frac{u}{c^2}\frac{\partial q}{\partial t} \right). \tag{3.9}$$

Note how the transformation of the velocity differs from that of a single relativistic particle under a boost ($\dot{q}' = (\dot{q} - u)/(1 - \dot{q}u/c^2)$).

To determine the Eulerian version of the transformation we invert the relation $x = q(a,t)$ to obtain $a(x,t)$. In fact, the mass function $m(x,t) = ka(x,t)$ can be expressed just in terms of the Eulerian hydrodynamic variables. To see this, we note that $\partial a/\partial x = (\partial q(a,t)/\partial a)^{-1}\big|_{a(x,t)} = \rho(x,t)/k$, i.e., $ka(x,t)$ is a mass-density potential. Using the continuity equation (2.2) we then have:

$$\frac{\partial a}{\partial t} = -\frac{\rho v}{k}, \quad \frac{\partial a}{\partial x} = \frac{\rho}{k}. \tag{3.10}$$

These relations are consistent with $a$ being a constant of the motion:

$$\frac{\partial a}{\partial t} + v\frac{\partial a}{\partial x} = 0. \tag{3.11}$$

This relation and the Bernoulli equation (2.13) form a potential system for the Chaplygin equations, the latter being obtained by differentiation of the former with respect to $x$. The transformation of the Eulerian fields $\rho, v$ follows from (3.9) and altogether we have the following symmetry of the Chaplygin equations:

$$\left. \begin{array}{l} x' = x, \quad t' = \gamma\left(t - \dfrac{u}{c}\dfrac{m(x,t)}{\sqrt{2\lambda}}\right), \quad m'(x',t') = \gamma\left(m(x,t) - \dfrac{u}{c}\sqrt{2\lambda}\,t\right), \\[1em] \dfrac{\sqrt{2\lambda}}{\rho'} = \gamma\left(\dfrac{\sqrt{2\lambda}}{\rho} + \dfrac{u}{c}v\right), \quad v' = \gamma\left(v + \dfrac{u}{c}\dfrac{\sqrt{2\lambda}}{\rho}\right). \end{array} \right\} \tag{3.12}$$

In these variables the Lorentz transformation has a hybrid relativistic/non-relativistic character. For example, the Eulerian deformation or 'speed' vector $(\sqrt{2\lambda}/\rho, v)$ ($\sqrt{2\lambda}/\rho$ is the local speed of sound) transforms as a covariant Lorentz 2-vector while the arguments of the fields transform asymmetrically.

The Lorentz transformation in this nonlinear realization thus comprises a field-dependent time transformation. Operationally, this represents a resynchronization of clocks with spacetime-dependent rates and settings. This generalizes the synchrony transformation customarily considered in special relativity ($x' = x, t' = t - \kappa x/c$ [17]) yet no compensating terms are required as the transformation is a symmetry of the physical equations that determine it. Consider a first Eulerian observer who monitors a Chaplygin flow at a point $x$ in a spatial reference frame equipped with a fleet of Einstein-synchronized clocks. A second observer, at rest with respect to the first, employs instead a clock at each point of the frame whose rate and setting are determined by $u$ and $m(x,t)$ according to the above formula for $t'$. Using this system she will also observe a Chaplygin flow but the density and velocity at each point are 'boosted' with respect to those recorded by the first observer at the same point. The parameter $u$ thus relates to variable clock rates and settings rather than to relatively moving frames. An additional Lagrangian perspective on $u$ is that, in accordance with the covariance group of (1.1), two observers monitoring a particular particle can utilize different labels $a, a'$ to identify it (the relativity of particle label). The respective labels are temporally invariable for each observer ($a'(x',t')$ obeys the primed version of (3.11)) and the observers register the same spatial coordinate at different times.

As noted above, (3.8) becomes a symmetry of (2.6) upon replacing $a$ by $k^{-1}\int \rho_0(a)da$.

The image in the $xt$-plane of the 'light cone' in the $at$-plane is the pair of curves $a(x_\pm(t),t) = \pm ct$. Differentiating with respect to $t$ and using (3.10), the gradients of the light-cone curves are given by

$$\frac{dx_\pm}{dt} = v \pm \frac{\sqrt{2\lambda}}{\rho}. \tag{3.13}$$

Recalling that $\sqrt{2\lambda}/\rho$ is the speed of sound for a Chaplygin fluid, these are the defining relations of the characteristic curves of a one-dimensional fluid [18]. Thus, *the pair of characteristics $x_\pm(t)$ of the Chaplygin fluid is the Eulerian image of the Lagrangian light cone $a = \pm ct$*, that is, $at$ provides a reference system with respect to which the characteristics are straight lines. Note that the gradients (3.13) are also the Riemann invariants $J_\mp$ of the fluid, which are conserved along the characteristics [18]. Using the notation $v_\pm = v \pm \sqrt{2\lambda}/\rho$ for the characteristic velocities, we have $J_\pm = v_\mp$ and the equations of motion (2.2) and (2.3) and the transformation of the fields in (3.12) may be expressed concisely as

$$\left(\frac{\partial}{\partial t} + v_\pm \frac{\partial}{\partial x}\right)v_\mp = 0, \quad v'_\pm = \left(\frac{c+u}{c-u}\right)^{\pm 1/2} v_\pm. \tag{3.14}$$

The Noether current corresponding to the Lorentz symmetry is given by the conserved angular momentum density,

$$m_{\sigma\mu\nu} = (1/c)(a_\sigma t_{\mu\nu} - a_\mu t_{\sigma\nu}), \quad \partial^\nu m_{\sigma\mu\nu} = 0, \tag{3.15}$$

from which the conserved charge is

$$M = (1/c)\int (a_0 t_{10} - a_1 t_{00})da = k\int \left[ct\frac{\partial q}{\partial t}\frac{\partial q}{\partial a} + a\frac{1}{2c}\left(\left(\frac{\partial q}{\partial t}\right)^2 + c^2\left(\frac{\partial q}{\partial a}\right)^2\right)\right]da \tag{3.16}$$

or, in the Eulerian variables,

$$M = \left(1/\sqrt{2\lambda}\right)\int \left(2\lambda tv + m(x,t)\mathfrak{H}\right)dx. \quad \text{(Lorentz transformation)} \tag{3.17}$$

This is the expression obtained by applying Noether's theorem in the Eulerian picture using the transformation (3.12).

### *3.3 Non-relativistic limits*

The fact that the Chaplygin fluid may be regarded as a relativistic system (i.e., the equations are covariant with respect to a realization of the Lorentz transformation) invites the question as to its nature when the conditions of the non-relativistic limit are obeyed. It is well known that a field equation can have more than one limiting form when $|u/c| \ll 1$, according to the relative magnitudes of the fields and their time variation [19]. In fact, in our case this multi-limit option is evident already in the kinematics as the transformation involves the field $m(x,t)$. There are four cases to consider:

(a) Assume $|m/t| \gg \sqrt{2\lambda}$ and $|v\rho| \gg \sqrt{2\lambda}$. Then (3.12), (2.2) and (2.3) reduce to

$$x' = x, \quad t' = t - \frac{u}{c}\frac{m(x,t)}{\sqrt{2\lambda}}, \quad \frac{\sqrt{2\lambda}}{\rho'} = \frac{\sqrt{2\lambda}}{\rho} + \frac{u}{c}v, \quad v' = v \tag{3.18}$$

$$\frac{\partial \rho}{\partial t} + \frac{\partial}{\partial x}(\rho v) = 0, \quad \frac{\partial v}{\partial t} + v\frac{\partial v}{\partial x} = 0. \tag{3.19}$$

(b) $|m/t| \gg \sqrt{2\lambda}$, $|v\rho| \ll \sqrt{2\lambda}$, $\partial \rho/\partial t = \partial v/\partial t = 0$:

$$x' = x, \quad t' = t - \frac{u}{c}\frac{m(x,t)}{\sqrt{2\lambda}}, \quad \rho' = \rho, \quad v' = v + \frac{u}{c}\frac{\sqrt{2\lambda}}{\rho} \tag{3.20}$$

$$\frac{\partial v}{\partial x} = \frac{\partial \rho}{\partial x} = 0. \tag{3.21}$$

(c) $|m/t| \ll \sqrt{2\lambda}$, $|v\rho| \ll \sqrt{2\lambda}$, $\partial \rho/\partial t = 0$:

$$x' = x, \quad t' = t, \quad \rho' = \rho, \quad v' = v + \frac{u}{c}\frac{\sqrt{2\lambda}}{\rho} \tag{3.22}$$

$$\frac{\partial}{\partial x}(\rho v) = 0, \quad \frac{\partial v}{\partial t} = \frac{\partial}{\partial x}\left(\frac{\lambda}{\rho^2}\right). \tag{3.23}$$

(d) $|m/t| \ll \sqrt{2\lambda}$, $|v\rho| \gg \sqrt{2\lambda}$, $\partial v/\partial t = 0$:

$$x' = x, \quad t' = t, \quad \frac{\sqrt{2\lambda}}{\rho'} = \frac{\sqrt{2\lambda}}{\rho} + \frac{u}{c}v, \quad v' = v \tag{3.24}$$

$$\frac{\partial \rho}{\partial t} + \frac{\partial}{\partial x}(\rho v) = 0, \quad \frac{\partial v}{\partial x} = 0. \tag{3.25}$$

In case (a) the 'non-relativistic' limit of the (non-relativistic!) Chaplygin gas is general free flow. The limits in the other cases are instances of this: in (b) and (c), $\rho, v$ are spacetime constants, and in (d), $\rho = \rho(x-vt,0), v = \text{const}$. In each case the set of limiting dynamical equations and the restrictions on the time dependence of $\rho$ and $v$ (in cases (b)-(d)) are exactly covariant with respect to the stated transformation. We note that in cases (a) and (b) the free flow admits a field-dependent time symmetry, which illustrates that this property is not specific to the Chaplygin gas (field-dependent symmetries are also known in point mechanics [20]).

With reference to case (b), consider a flow for which $\rho = k, v = 0$. Then, from (3.10), $a = x$ (we ignore the additive constant in $a$, which corresponds to a time translation) and the kinematical transformation (3.20) reduces to $x' = x, t' = t - (u/c^2)x$ with $|x/ct| \gg 1$. This transformation is an element of what Lévy-Leblond called the 'Carroll' group, a limiting case of the Lorentz transformation complementary to the Galileo boost [21,22]. We see from (3.12) that the relativistic antecedent of the Carroll group need not be a Lorentz transformation in the same independent variables.

## 4 Galileo symmetry

We can generate the space translation and boost components of the Galileo group by applying the superposition principle in the Lagrangian picture; (1.1) is covariant with respect to the following transformation:

$$q'(a',t') = q(a,t) - wt + s, \quad t' = t, \quad a' = a, \qquad (4.1)$$

where $s, w$ are constants. The corresponding Eulerian relations are likewise those of a Galileo boost and space translation:

$$x' = x - wt + s, \quad t' = t, \quad \rho'(x',t') = \rho(x,t), \quad v'(x',t') = v(x,t) - w. \qquad (4.2)$$

The boost describes the relation between the coordinates employed by two relatively moving observers. Wearing their Lagrangian hat, the observers ascribe the coordinates to the same particle. In their Eulerian guise, they ascribe them to the same space point at which they monitor the same set of particles.

The time translation component of the Galileo group has been studied in Sec. 3.1. The remaining Noether charges in the two pictures are

$$P = k \int \frac{\partial q}{\partial t} da = \int \rho v \, dx \quad \text{(space translation)} \qquad (4.3)$$

$$B = k \int \left( q - t \frac{\partial q}{\partial t} \right) da = \int \rho(x - tv) dx. \quad \text{(Galileo boost)} \qquad (4.4)$$

In all cases the Lagrangian-picture conserved charges written in the Eulerian variables coincide with those obtained by applying Noether's theorem directly in the Eulerian picture.

## 5 Other kinematical symmetries

We mention two other obvious symmetries of the Chaplygin system. The first, apparently not known, derives from the two-parameter ($\varepsilon^\mu$) special conformal symmetry of the wave equation whose infinitesimal form is

$$q' = q, \quad a'^\mu = a^\mu + \varepsilon^\sigma c^{-1}\left(\delta^\mu_\sigma a_\nu a^\nu - 2a^\mu a_\sigma\right). \tag{5.1}$$

The implied infinitesimal field-dependent symmetry is

$$\left.\begin{array}{l} x' = x, \quad t' = t - \varepsilon^0\left(t^2 + m^2/2\lambda\right) + 2\varepsilon^1 t m/\sqrt{2\lambda}, \\ \rho' = \rho\left[1 - 2\varepsilon^0\left(t + \rho v m/2\lambda\right) + 2\varepsilon^1\left(t\rho v + m\right)/\sqrt{2\lambda}\right], \\ v' = v + 2\varepsilon^0\left(tv + m/\rho\right) - 2\varepsilon^1\sqrt{2\lambda}\left(v m/2\lambda + t/\rho\right). \end{array}\right\} \tag{5.2}$$

The Noether charges implied by the two transformations (5.1) become in the Eulerian picture:

$$\left.\begin{array}{l} S_{\varepsilon^0} = (c/2\lambda)\int\left[\mathfrak{H}\left(2\lambda t^2 + m^2\right) + 4\lambda tvm\right]dx \\ S_{\varepsilon^1} = -(1/k)\int\left[2tm\mathfrak{H} + v\left(2\lambda t^2 + m^2\right)\right]dx. \end{array}\right\} \text{(special conformal)} \tag{5.3}$$

The second transformation is a known Chaplygin symmetry [11], a dilation: $q' = q, t' = dt, a' = da$ whence $x' = x, t' = dt, \rho' = d\rho, v' = d^{-1}v$ with corresponding charge $D = \int(t\mathfrak{H} - \rho\theta)dx$. A dilation of the field, $q' = fq$, is an instance of the superposition symmetry studied in Sec. 7.

## 6 The potentials as solutions and their transformations

It is easy to check that each of the Lagrangian-picture potentials $\phi$, $\alpha$ and $\beta$ in Table 1 is a solution to (1.1) (denoted $\bar{q}$) if $q$ is. Introducing a parameter $\omega$ having the dimension of inverse speed, the potentials generate the following new Chaplygin solutions from a given solution $\rho, v$, via the formulas (2.1) (with $q$ replaced by $\bar{q}$):

$$\left.\begin{array}{l} \bar{q} = \omega\phi : \rho_\phi = \rho/\omega v, \quad v_\phi = \omega\mathfrak{H}/\rho \\ \bar{q} = \omega\alpha : \rho_\alpha = k/\omega v, \quad v_\alpha = 2\lambda\omega/k\rho \\ \bar{q} = \omega^2\beta : \rho_\beta = k\rho/\omega^2\mathfrak{H}, \quad v_\beta = 2\lambda\omega^2 v/k\rho. \end{array}\right\} \tag{6.1}$$

Of these solutions, $\alpha$ is a Lorentz scalar and the couple $(c\phi, \beta)$ transforms as a covariant 2-vector. The complete set of transformations of the Eulerian potentials under a Lorentz transformation (3.12) is then

$$\xi'(x',t') = \xi(x,t), \quad \theta' = \gamma\left(\theta + \frac{u}{c\sqrt{2\lambda}}\chi\right), \quad \chi' = \gamma\left(\chi + \frac{u}{c}\sqrt{2\lambda}\theta\right),$$
$$m' = \gamma\left(m - \frac{u}{c}\sqrt{2\lambda}\,t\right) \quad (6.2)$$

where the arguments transform as $x' = x, t' = \gamma\left(t - um(x,t)/c\sqrt{2\lambda}\right)$. The potential equations (2.13) and (3.11) are correspondingly Lorentz covariant.

## 7 Superposition symmetries

### 7.1 Nonlinear superposition rule for the Chaplygin equations

The linearization (1.1) of the Chaplygin equations provides a simple way to generate a (nonlinear) superposition principle for the Eulerian fields from the linear rule of combination supported by the Lagrangian trajectories. The symmetries of Sec. 4 involve application of linear superposition to establish the covariance of the wave equation. Generalizing, we may regard the linear superposition operation, which builds a new solution from a given one, as a form of symmetry operation with respect to which the wave equation is covariant. This is an infinite dimensional Abelian group characterized by a transformation function that solves the dynamical equation. Noether's theorem generates an associated constant of the motion.

Consider two solutions $q(a,t)$, $\bar{q}(a,t)$ of (1.1). Linearly superposing them supplies a new solution, which we express as a transformation:

$$q'(a',t') = q(a,t) + \bar{q}(a,t), \quad t' = t, \quad a' = a. \quad (7.1)$$

Let the Eulerian functions corresponding to the solution $\bar{q}(a,t)$ be denoted $\bar{\rho}(x,t), \bar{v}(x,t)$, which may be computed from (2.1). Then the two Chaplygin solutions $(\rho,v), (\bar{\rho},\bar{v})$ combine to generate a third solution according to the following scheme:

$$x' = x + \bar{x}(x,t), \quad t' = t, \quad \rho'(x',t') = \frac{\rho(x)\bar{\rho}(\bar{x}(x))}{\rho(x) + \bar{\rho}(\bar{x}(x))}, \quad v'(x',t') = v(x) + \bar{v}(\bar{x}(x)) \quad (7.2)$$

where

$$\bar{x}(x,t) = \bar{q}(a(x,t),t). \quad (7.3)$$

The transformation (7.2) is a nonlinear symmetry of the Chaplygin theory.

We can apply this method to obtain a new solution to the Chaplygin equations from two known ones $(\rho,v), (\bar{\rho},\bar{v})$ in two ways:

(a) *Using trajectories.* Solve $v = \partial q/\partial t$ and $\bar{v} = \partial \bar{q}/\partial t$ to obtain two sets of trajectories $q, \bar{q}$. Create a new set $q'$ using (7.1). Derive the new solution $\rho', v'$ by differentiation and inversion via (2.1).

(b) *Without using trajectories*. Differentiate (7.3) with respect to $x$ and $t$ and use the expressions (3.10) for $a(x,t)$ to obtain:

$$\frac{\partial \bar{x}(x,t)}{\partial x} = \frac{\rho(x,t)}{\bar{\rho}(\bar{x}(x,t),t)}, \quad \frac{\partial \bar{x}(x,t)}{\partial t} = \bar{v}(\bar{x}(x,t),t) - \frac{\rho(x,t)v(x,t)}{\bar{\rho}(\bar{x}(x,t),t)}. \tag{7.4}$$

Solve these differential equations, invert the first member of (7.2) to obtain $x(x',t)$ and insert the latter in the right-hand sides of the last two members.

In the examples below we consider cases where the function (7.3) can be expressed in terms of the hydrodynamic potentials associated with $\rho,v$ (and linear functions of $t$). As a first example, suppose $\bar{q}(a,t) = (u/c)a$. Then

$$x' = x + u\frac{m(x,t)}{\sqrt{2\lambda}}, \quad t' = t, \quad \frac{1}{\rho'} = \frac{1}{\rho} + \frac{u}{\sqrt{2\lambda}}, \quad v' = v, \tag{7.5}$$

which is a field-dependent coordinate symmetry complementary to the time transformation (3.12).

The Noether charge associated with the linear superposition symmetry (7.1) is the scalar product of the two solutions:

$$S = k\int \left( \bar{q}\frac{\partial q}{\partial t} - q\frac{\partial \bar{q}}{\partial t} \right) da. \tag{7.6}$$

The constants of the motion (4.3) and (4.4) are included in this formula as obvious special cases (let $\bar{q} = s, -wt$). In the Eulerian picture we therefore have the following conserved quantity:

$$S = \int \left[ \bar{x}(x,t)v(x,t) - x\bar{v}(\bar{x}(x,t),t) \right] \rho dx. \tag{7.7}$$

### 7.2 Deconstruction of spacetime mixing transformation

A wide variety of previously unknown field-dependent Chaplygin symmetries may be obtained by employing the potentials $\phi(a,t), \alpha(a,t), \beta(a,t)$ as components of transformations applied to the wave equation. As an example, we derive the spacetime mixing symmetry given previously [11] by combining two independent field-dependent transformations (one of space, one of time). In the following $\omega_1$ and $\omega_2$ are constants having the dimension of inverse speed.

First, consider the following superposition transformation involving the velocity potential:

$$q'(a',t') = q(a,t) + \omega_1\phi(a,t), \quad t' = t, \quad a' = a, \tag{7.8}$$

which, following Sec. 6, is a solution. The corresponding Eulerian symmetry is

$$x' = x + \omega_1 \theta(x,t), \quad t' = t, \quad \rho' = \frac{\rho}{1+\omega_1 v}, \quad v' = v + \omega_1\left(\tfrac{1}{2}v^2 + \frac{\lambda}{\rho^2}\right). \tag{7.9}$$

Second, we observe that the following field-dependent transformation of the Lagrangian variables is a symmetry of (1.1):

$$\left.\begin{aligned} q'(a',t') &= q(a,t), \quad t' = t + \omega_2 q(a,t) + \tfrac{1}{2}\omega_2^2 \phi(a,t), \\ a' &= a + \omega_2 \alpha(a,t) + \tfrac{1}{2}\omega_2^2 \beta(a,t). \end{aligned}\right\} \tag{7.10}$$

The corresponding transformation of the Eulerian fields may be expressed compactly using the characteristic velocities $v_\pm = v \pm \sqrt{2\lambda}/\rho$:

$$x' = x, \quad t' = t + \omega_2 x + \tfrac{1}{2}\omega_2^2 \theta(x,t), \quad v'_\pm = \frac{v_\pm}{\left(1+\tfrac{1}{2}\omega_2 v_\pm\right)^2}. \tag{7.11}$$

The two transformations just stated, the superposition and the label-time, have the same associated Noether charge[1]:

$$\left.\begin{aligned} G &= k \int \left(\phi \frac{\partial q}{\partial t} - q\tfrac{1}{2}\left[c^2\left(\frac{\partial q}{\partial a}\right)^2 + \left(\frac{\partial q}{\partial t}\right)^2\right]\right) da \\ &= \int (\theta \rho v - x \mathfrak{H}) dx. \end{aligned}\right\} \text{(superposition/label-time mixing)} \tag{7.12}$$

Combining the transformations (7.8) and (7.10) we have the following 2-parameter symmetry:

$$\left.\begin{aligned} q'(a',t') &= q(a,t) + \omega_1 \phi(a,t), \quad t' = t + \omega_2 q(a,t) + \tfrac{1}{2}\omega_2^2 \phi(a,t), \\ a' &= a + \omega_2 \alpha(a,t) + \tfrac{1}{2}\omega_2^2 \beta(a,t) \end{aligned}\right\} \tag{7.13}$$

or

$$x' = x + \omega_1 \theta(x,t), \quad t' = t + \omega_2 x + \tfrac{1}{2}\omega_2^2 \theta(x,t), \quad v'_\pm = \frac{v_\pm\left(1+\tfrac{1}{2}\omega_1 v_\pm\right)}{\left(1+\tfrac{1}{2}\omega_2 v_\pm\right)^2}. \tag{7.14}$$

Setting $\omega_1 = \omega_2 = \omega$ we obtain finally the desired 1+1-dimensional version of the spacetime mixing transformation:

$$\left.\begin{aligned} x' &= x + \omega \theta(x,t), \quad t' = t + \omega x + \tfrac{1}{2}\omega^2 \theta(x,t), \\ \rho' &= \rho\left(1 + \omega v - \tfrac{1}{2}\omega^2 \partial\theta/\partial t\right), \quad v' = \frac{v - \omega \partial\theta/\partial t}{1 + \omega v - \tfrac{1}{2}\omega^2 \partial\theta/\partial t}. \end{aligned}\right\} \tag{7.15}$$

---

[1] This provides an example of how Noether's theorem establishes just a correlation between a symmetry and a constant of the motion rather than a unique correspondence, even for a given Lagrangian.

The associated constant of motion is, of course, (7.12).
    As another example of a field-dependent transformation, we can interchange $ct$ and $a$ in (7.10) to obtain a Chaplygin symmetry involving the momentum density and energy density potentials:

$$x' = x, \quad ct' = ct + \omega_2 \xi(x,t) + \tfrac{1}{2}\omega_2^2 \chi(x,t), \quad \tilde{v}'_\pm = \frac{\tilde{v}_\pm}{\left(1 + \tfrac{1}{2}\omega_2 \tilde{v}_\pm\right)^2} \qquad (7.16)$$

with $\tilde{v}_\pm = \pm v + \sqrt{2\lambda}/\rho$. The implied Noether charge is

$$\bar{D} = \left(1/\sqrt{2\lambda}\right)\int (2\lambda\theta - \xi\mathfrak{H})\,dx. \quad \text{(dual label-time mixing)} \qquad (7.17)$$

### 7.3 Connection with Born-Infeld theory

As a final application of the superposition symmetry we observe that the Lagrangian representation of the kinematical Lorentz symmetry exhibited by the Born-Infeld equations, which are obtained from the Chaplygin set by a field-dependent change of variables, is provided by a slight modification of the 2-parameter symmetry (7.13). Starting from the Chaplygin equations, define the following set of variables [9]:

$$x_B = x, \quad t_B = \left(t + \theta/c^2\right)/\sqrt{2}, \quad \theta_B = \left(c^2 t - \theta\right)/\sqrt{2}. \qquad (7.18)$$

The (Born-Infeld) field equations written in these variables are covariant with respect to a Lorentz transformation:

$$x'_B = \gamma(x_B - u t_B), \quad t'_B = \gamma(t_B - u x_B/c^2), \quad \theta'_B = \theta_B. \qquad (7.19)$$

Combining (7.18) and (7.19) it is easy to show that the corresponding Lagrangian symmetry is obtained from (7.13) as follows: let $\omega_1 = -u/\sqrt{2}c^2$, $\omega_2 = -\sqrt{2}\gamma u/c^2(\gamma+1)$ with $\gamma = \left(1 - u^2/c^2\right)^{-1/2}$. Multiply the right-hand sides of $t'$, $a'$ and $q'$ by $\tfrac{1}{2}(\gamma+1)$, $\tfrac{1}{2}(\gamma+1)$ and $\gamma$, respectively. Finally, add $-\gamma u t/\sqrt{2}$ to the modified $q'$.

### 8 Connection with quantum mechanics

Consider in one-dimensional fluid mechanics a steady isentropic flow (with equation of state $p = p(\rho)$) subject to an external force. The continuity equation implies that $\rho v = K = \text{const.}$ (assumed to be non-zero) and the total force per unit mass according to Euler's equation is therefore

$$-\frac{d}{dx}\left(\bar{V} + \int \frac{dp}{\rho}\right) = \frac{d}{dx}\left(\frac{K^2}{2\rho^2}\right) \qquad (8.1)$$

where $\bar{V}$ is the external potential per unit mass. Comparing with (2.3) and identifying $2\lambda$ with $K^2$, we deduce that a finite-current steady isentropic one-dimensional flow subject to an external force behaves as a Chaplygin gas. Integrating with respect to $x$ and substituting for $p(\rho)$, (8.1) supplies a formula for $\rho$.

This result has currency beyond the classical domain; it may be applied to the hydrodynamic representation of a one-dimensional quantum mechanical system (of mass $m$ in a potential $V$). It has been shown by the author that quantal evolution may be computed independently of the wavefunction by representing the quantum state as the collective motion $q(a,t)$ of a continuum of particles (whose initial conditions are correlated with the initial wavefunction). Introducing an equation of state descriptive of quantum systems, this quantum Lagrangian picture provides a formula for the time dependence of the wavefunction in terms of the trajectories. In one dimension the Lagrangian is [15,23]

$$L = \int \bar{k} \left\{ \frac{1}{2} m \left( \frac{\partial q}{\partial t} \right)^2 - \frac{\hbar^2}{8m} \left[ \frac{\partial}{\partial a} \left( \frac{\partial q}{\partial a} \right)^{-1} \right]^2 - V(q) \right\} da \qquad (8.2)$$

where we choose a labelling in which the density $k = m\bar{k}$ is uniform. The (nonlinear) Euler-Lagrange equation for $q$ implies the function $\psi(x,t) = \sqrt{\rho} \exp(i\theta/\hbar)$ obeys the Schrödinger equation in the Eulerian picture. In the case of a stationary state of energy $E$, the Schrödinger equation implies $\rho v = K$ and the Ermakov equation [24]:

$$\hbar^2 \frac{d^2 \rho^{1/2}}{dx^2} + 2m(E-V)\rho^{1/2} = m^2 K^2 \rho^{-3/2}. \qquad (8.3)$$

The latter, which determines $\rho$, implies (8.1) with $\bar{V} = V/m$ and $p = -(\hbar^2 \rho / 4m^2) d^2 \log \rho / dx^2$ (the quantum equation of state). Thus, *a one-dimensional quantum system in a stationary state behaves as a Chaplygin gas*.

Under a Lorentz transformation (3.12) the time-independence of $\rho, v$ is preserved and the relation $\rho v = \sqrt{2\lambda}$ is covariant. Eq. (8.1) is covariant when $\bar{V} = 0$ and $p$ is the Chaplygin expression (when (8.1) reduces to an identity). The Lorentz transformation is not a symmetry of the Schrödinger equation.

### 9 Higher dimensions

#### *9.1 Wave equation*

The extension to higher dimensions involves consideration of the dimension of both the medium and the embedding space. If we consider an elastic string in three dimensions, each spatial coordinate $q^i$, $i = 1,2,3$, obeys the 1+1 equation (1.1) where $a$ is a coordinate along the string [25] (note that the results of this section hold in any space dimension $d$). On the other hand, for a medium that supports longitudinal and transverse vibrations,

$$\frac{\partial^2 q^i}{\partial t^2} - c^2 \frac{\partial^2 q^i}{\partial a^j \partial a^j} = -J^i(a,t), \qquad (9.1)$$

where we have included a source term. In mapping this equation into the Eulerian picture the equation of state involves individual elements of the deformation matrix $\partial q_i/\partial a_j$ and cannot be written just in terms of its determinant, which indicates that the medium is not assimilable to a fluid system, and in particular not to the Chaplygin gas. Nevertheless, the theory still exhibits a dual relativistic/non-relativistic character: it is Poincaré covariant in its Lagrangian version if $q^i(a,t)$ form a triplet of scalar fields, and Galileo covariant in its Eulerian version if the Lagrangian field is a vector:

$$q'^i = R^i_j q^j - w^i t + s^i, \quad t' = t + t_0, \quad a'^i = a^i. \qquad (9.2)$$

It is assumed the source transforms appropriately in each case. As in the 1+1 case, the Eulerian image of the Lagrangian picture Lorentz transformation involves a field-dependent transformation of the time:

$$x'^i = x^i, \quad t' = \gamma\left(t - \frac{u^i a^i(x,t)}{c^2}\right), \quad a'^i(x',t') = \left[\frac{(\gamma-1)}{u^k u^k} u^j u^i + \delta^i_j\right] a^j(x,t) - \gamma u^i t. \quad (9.3)$$

This represents a resynchronization 'boost' as described in Sec. 3.2 while a spatial boost is implemented by the Galileo transformation.

### *9.2 Maxwell's equations*

It is instructive to employ this continuum-mechanical representation to model Maxwell's equations, where the dual symmetries are executed by transformation rules for $q^i$ partly different to those just stated. In this case a spatial boost is implemented by a Lorentz transformation and the Galileo boost has a formal character.

Maxwell's equations are obtained if we append to (9.1) the equations [26]

$$\frac{\partial^2 q^0}{\partial a^j \partial a^j} = -P(a,t), \quad \frac{\partial q^i}{\partial a^i} = 0, \quad \text{(Coulomb gauge)} \qquad (9.4)$$

and identify the source $J^i$ with the transverse current, $q^i$ with the magnetic vector potential, and $q^0$ with the Coulomb potential whose source is the charge density $P$. The sources are prescribed functions (connected by the continuity equation). This formalism is covariant with respect to a gauge-dependent Poincaré transformation in *at* space whose infinitesimal form is

$$q'^\mu(a',t') = q^\mu(a,t) - \varepsilon^\mu{}_\nu q^\nu(a,t) + \eta^{\mu\nu} \partial \Lambda/a^\nu, \quad a'^\mu = a^\mu - \varepsilon^\mu{}_\nu a^\nu + \varepsilon^\mu \qquad (9.5)$$

where $\mu,\nu = 0,1,2,3$, $a^\mu = (ct, a^i)$, $\varepsilon_{\mu\nu} = -\varepsilon_{\nu\mu}$, $\varepsilon^0{}_i = u^i/c$ represents a boost, $\varepsilon^i{}_j$ a 3-rotation, $\varepsilon^\mu$ a translation, $\eta^{\mu\nu} = (1, -\delta_{ij})$, and the gauge function depends on the boost parameter via the relation $\partial^2 \Lambda / \partial a^i \partial a^i = (u^i/c^2)(\partial q^i/\partial t - c \partial q^0/\partial a^i)$.

Passing to the independent variables $x^i (= q^i), t$, the dependent variables are $v^i(x,t) = \partial q^i/\partial t |_{a(x,t)}$, $a^i(x,t)$, and $Q(x,t) = q^0(a(x,t), t)$. Denoting the deformation matrix by $D^i{}_j(x,t) = (\partial a^i/\partial x^j)^{-1}$, the Maxwell equation (9.1) translates into an Euler-type equation,

$$\frac{\partial v^i}{\partial t} + v^j \frac{\partial v^i}{\partial x^j} = c^2 D^k{}_j \frac{\partial D^i{}_j}{\partial x^k} - J^i(a(x,t), t), \qquad (9.6)$$

which, as noted above, involves explicit reference to the deformation matrix[2]. To obtain a closed set of equations in the dependent variables we need the Eulerian versions of (9.4), namely,

$$D^k{}_i \frac{\partial}{\partial x^k}\left(D^j{}_i \frac{\partial Q}{\partial x^j}\right) = -P(a(x,t), t), \quad D^i{}_i = D^j{}_i \frac{\partial v^i}{\partial x^j} = 0, \qquad (9.7)$$

and the equation of motion of the particle label, which is the condition that it is conserved by the flow:

$$\frac{\partial a^i}{\partial t} + v^j \frac{\partial a^i}{\partial x^j} = 0. \qquad (9.8)$$

Eqs. (9.6)-(9.8) are Maxwell's equations in the new variables. The density $\rho(x,t) = kD^{-1}$, $D = \det D^i{}_j$, may be computed from the solution.

The Poincaré group takes the following form in the Eulerian description. An infinitesimal pure gauge-dependent Lorentz transformation (9.5) becomes the following field-dependent transformation:

$$\left.\begin{aligned}
x'^i &= x^i - u^i t + D^{ji} \frac{\partial \lambda}{\partial x_j}, \quad t' = t - \frac{u^i}{c^2} a^i(x,t), \quad a'^i(x', t') = a^i(x,t) - u^i t, \\
v'^i(x', t') &= v^i(x,t) - \frac{u^i}{c}\left(\frac{\partial Q}{\partial t} - D^i{}_j \frac{\partial Q}{\partial x^i} \frac{\partial a^j}{\partial t}\right) + u^j D^i{}_j \\
&\quad - D^k{}_i \frac{\partial}{\partial x^k}\left(\frac{\partial \lambda}{\partial t} - D^l{}_j \frac{\partial \lambda}{\partial x^l} \frac{\partial a^j}{\partial t}\right), \\
Q'(x', t') &= Q(x,t) - \frac{u^i}{c} a^i(x,t) + \frac{\partial \lambda}{c \partial t} - D^i{}_j \frac{\partial \lambda}{\partial x^i} \frac{\partial a^j}{c \partial t},
\end{aligned}\right\} \qquad (9.9)$$

---

[2] An alternative fluid-mechanical model of Maxwell's equations in which the Euler force can be written just in terms of the density and its derivatives via a quantum potential-like term has been developed elsewhere [22].

with $\lambda(x,t) = \Lambda(a(x,t),t)$ and $D^k{}_i \partial(D^j{}_i \partial\lambda/\partial x^j)/\partial x^k = (u^i/c^2)(v^i - cD^j{}_i \partial Q/\partial x^j)$.
An *at* 3-rotation and label-time translation corresponds to an *xt* 3-rotation and label-time translation:

$$\left.\begin{array}{l} x'^i = x^i - \varepsilon^i_j x^j, \quad t' = t + \varepsilon^0, \quad a'^i(x',t') = a^i(x,t) - \varepsilon^i_j a^j(x,t) + \varepsilon^i, \\ v'^i(x',t') = v^i(x,t) - \varepsilon^i_j v^j(x,t), \quad Q'(x',t') = Q(x,t). \end{array}\right\} \quad (9.10)$$

The covariance of the Eulerian theory with respect to a Galileo boost and a coordinate translation may be obtained as special cases of the field-dependent symmetry corresponding to the residual gauge freedom in the Lagrangian picture, namely, $q'^\mu = q^\mu + \partial\Lambda/\partial a_\mu, a'^\mu = a^\mu$ with $\partial^2 \Lambda/\partial a^i \partial a_i = 0$. The formula for the corresponding Eulerian transformation is obtained by setting $u^i = 0$ in (9.9), subject to $D^k{}_i \partial(D^j{}_i \partial\lambda/\partial x^j)/\partial x^k = 0$. A combined Galileo boost and coordinate translation corresponds to the choice $\Lambda = a_i(-w^i t + s^i)$:

$$\left.\begin{array}{l} x'^i = x^i - w^i t + s^i, \quad t' = t, \quad a'^i(x',t') = a^i(x,t), \quad v'^i(x',t') = v^i(x,t) - w^i, \\ Q'(x',t') = Q(x,t) - \dfrac{w^i}{c} a^i(x,t). \end{array}\right\} \quad (9.11)$$

We have thus demonstrated that Maxwell's equations admit both Poincaré and Galileo covariant formulations depending on the independent variables adopted. The Galileo label-time translation and 3-rotation correspond to the analogous components of the Poincaré group, the Galileo coordinate translation and boost correspond to elements of the pure gauge group, and a field-dependent *xt* symmetry corresponds to the pure gauge-dependent Lorentz transformation. Of course, interpreting the theory in the usual way as comprising statements about the time variation of fields at fixed space points (so that our Lagrangian picture is treated as an Eulerian picture), what we have termed the Eulerian picture becomes an abstract ('super-Eulerian') representation where points $x^i$ are labelled by the components of the vector potential and the electromagnetic field is represented by the fields $v^i, a^i, Q$ on this space. The Galileo covariance does not therefore relate to spatial reference frames. Indeed, we may regard a gauge transformation obeying Laplace's equation as a generalized Galileo boost.

It has been shown elsewhere that the Lagrangian-picture theory corresponding to the wave equation regarded as a statement in an Eulerian picture involves a velocity-dependent force and the Eulerian Lorentz transformation is mirrored by a label-dependent Lagrangian Lorentz transformation [27].

### *9.3 Chaplygin gas*

Given that the Lagrangian-picture linear wave equation in more than one dimension does not map into the higher-dimensional Chaplygin gas, we might ask what does. To obtain the Lagrangian picture version, we start from the three-dimensional Chaplygin Euler equation,

$$\frac{\partial v^i}{\partial t} + v^j \frac{\partial v^i}{\partial x^j} = \frac{\partial}{\partial x^i}\left(\frac{\lambda}{\rho^2}\right) \tag{9.12}$$

and write its solutions in terms of the trajectories using the conversion formulas

$$\rho(x(a,t),t) = kJ^{-1}, \quad v^i(x(a,t),t) = \frac{\partial q^i(a,t)}{\partial t}, \quad J = \det \partial q^i/\partial a^j. \tag{9.13}$$

Noting that the right-hand side of (9.12) may be written

$$\tfrac{1}{2}c^2 \frac{\partial J^2}{\partial q^i} = c^2 J \frac{\partial}{\partial q^j}\left(J \frac{\partial q^i}{\partial q^j}\right) \tag{9.14}$$

and using the cofactor $J_i^{\;j} = \partial J/\partial(\partial q^i/\partial a^j)$ we obtain the following nonlinear equation:

$$\frac{\partial^2 q^i}{\partial t^2} - c^2 J_l^{\;j} \frac{\partial}{\partial a^j}\left(J_l^{\;k} \frac{\partial q^i}{\partial a^k}\right) = 0. \tag{9.15}$$

This equation has been derived by Bazeia from the Nambu-Goto brane action [10]. The Lorentz covariance of the 1+1 theory in the *at* variables is evidently broken and it may be shown that there do not exist independent variables $t'(a,t), a'^i(a,t)$ for which (9.15) takes the d'Alembertian form. The Lorentz covariance of the Chaplygin gas regarded as an image of the wave equation is therefore an artefact of the one-dimensional treatment, as is the superposition symmetry (and hence the decomposition of the mixing symmetry) in Sec. 7. Eq. (9.15) admits (9.2) as a symmetry, whence we confirm the Galileo covariance of the higher-dimensional Chaplygin gas.

## 10 Conclusion

We have sought to support the claim that it can be meaningful to assert the simultaneous Poincaré and Galileo covariance of a system of equations. No contradiction arises as the symmetries apply to distinct sets of independent variables (of equal dimension in the cases we consider) and the transformation parameters have correspondingly distinct meanings. The physical significance to be attributed to each symmetry depends on the interpretation of the independent variables, which need not involve spatial coordinates. In the case of the wave equation where the field is a displacement function both parameters have physical significance, to do with resynchronization and spatial boost. On the other hand, in the context of Maxwell's equations, one parameter represents a boost and the other a gauge transformation. Whether one regards the simultaneous manifestation of the Poincaré and Galileo symmetries as conceptually significant or not, we have shown that this structure may be exploited to reveal novel characteristics about a system. Remarkably, this is true even for the one-dimensional wave equation whose simple properties may be utilized to discover hitherto unnoticed features of the non-relativistic Chaplygin gas including a relativistic structure with a limiting case that exhibits the Carroll group, novel field-dependent symmetries and associated Noether charges, and a nonlinear superposition principle.

In this endeavour we have advocated employing continuum physics as a mathematical and conceptual resource since its principal pictures – the Eulerian and the Lagrangian – provide natural complementary settings for the actions of the two groups. In particular, a relabelling diffeomorphism is available to simplify the Lagrangian description and thus provide simple models for possibly complex Eulerian formulations. One may in any case always simulate a field system by regarding it as a Lagrangian system, as we have done with Maxwell's equations. In fact, a hierarchy of Eulerian and Lagrangian models may be generated from a set of field equations by a succession of mappings from one framework to the other. The interchange of independent and dependent variables intrinsic to the continuum-physical approach implies that the appearance of field-dependent symmetries is likely to be a generic feature for any field equation presented in this language. Along these lines a source of further examples in higher dimensions could be provided by brane models where brane transformations induce non-trivial world coordinate transformations. Naturally, similar results may be expected for other groups linked by asymptotic inclusion.